\documentclass[onecolumn,showpacs,preprintnumbers,amsmath,amssymb]{revtex4}
\usepackage{graphicx}
\usepackage{bm}
\usepackage{dcolumn}

\def\prl#1#2#3{{ Phys.   Rev.   Lett.  } {\bf #1}, #2 (#3)}
\def\pla#1#2#3{Phys.   Lett.   A {\bf #1}, #2 (#3)}

\def\pre#1#2#3{Phys.   Rev.   E {\bf #1}, #2 (#3)}

\def\jpa#1#2#3{J.   Phys.   A {\bf #1}, #2 (#3)}

\def\rmp#1#2#3{Rev.   Mod.   Phys.   {\bf #1}, #2 (#3)}

\def\H{{\cal H}}
\def\Hp{H_1}
\def\eff{\mathrm{eff}}

\def\noi{\noindent}
\def\bc{\begin{center}}
\def\ec{\end{center}}
 \newcommand{\bea}{\begin{equation}}
 \newcommand{\eea}{\end{equation}\noi}
 \newcommand{\ber}{\begin{eqnarray}}
 \newcommand{\eer}{\end{eqnarray}\noi}
\begin{document}
\title{Approach to equilibrium in adiabatically evolving potentials}
\author{H.   S.   Samanta}\email{tphss@mahendra.iacs.res.in}
\author{J.   K. Bhattacharjee}\email{tpjkb@mahendra.iacs.res.in}
\affiliation{Department of Theoretical Physics,
Indian Association for the Cultivation of Science \\
Jadavpur, Calcutta 700 032, India}
\author{R.Ramaswamy}\email{r.ramaswamy@mail.jnu.ac.in} \affiliation{ School
of Physical Sciences, Jawaharlal Nehru University, New Delhi 110
067, India}
\date{\today}
\begin{abstract}
For a potential function (in one--dimension) which evolves 
from a specified initial form, $V_i(x)$,  to a different 
$V_f(x)$ asymptotically, we study the evolution, 
in an overdamped dynamics, of an 
initial probability density to its final equilibrium.  
There can be unexpected effects that can arise from 
the time dependence. We choose a time variation of the form
$V(x,t)=V_{f}(x)+(V_{i}-V_{f})e^{-\lambda t}$. For a $V_{f}(x)$, 
which is double welled and a $V_{i}(x)$ which is simple harmonic,  
we show that in particular, if the evolution is adiabatic, 
this results in a {\it decrease} in the Kramers' time characteristic 
of $V_f(x)$. Thus the time dependence makes diffusion over 
a barrier more efficient. There can also be interesting resonance 
effects when $V_{i}(x)$ and $V_{f}(x)$ are two harmonic 
potentials displaced with respect to each other
that arise from the coincidence of the intrinsic 
time--scale characterizing the potential variation and the 
Kramers' time. Both these features are illustrated through 
representative examples.
\end{abstract}

\maketitle

\section{Introduction}
The pioneering work of Kramers \cite{1} on thermally activated
barrier crossing has provided an understanding of the microscopic
mechanism underlying the Arrhenius temperature dependence of
crossing rates \cite{2}. Several variants of the basic problem have
subsequently been studied in the literature. A fair amount of
attention has recently been devoted to the study of more complex
nonequilibrium systems. These include the cases of  diffusion over
a barrier in the presence of harmonic force \cite{3,4,5,6,7,8,9,10}
and diffusion over a fluctuating barrier \cite{11,12,13,14,15,16,17}.
The hallmark of the former situation is the  phenomenon of
stochastic resonance , where the signal--to--noise ratio of the
system response to an applied harmonic force displays a local
maximum as a function of the diffusion constant or the
temperature. In fluctuating barriers,  the discovery \cite{12} that
the mean first passage time has a minimum as a function of the
correlation time characterizing the fluctuation has prompted a
wide variety of investigations. The problem of surmounting
potential barriers \cite{17,18,19} has gained importance in other fields
of science  such as evolutionary computations \cite{20,21} and
global optimization \cite{22} as well.

In the present work we consider the situation of barrier crossing
of a time--dependent potential which adiabatically evolves from
$V_{i}(x)$ at $t=0$ to the potential $V_{f}(x)$ as $t \to \infty
$.   In such a situation there will be an eventual equilibrium
distribution given by \bea
P_{eq}\sim\exp{-\frac{V_{f}(x)}{\epsilon}}, \eea where $\epsilon$
is the diffusion constant and the approach to this equilibrium
will be governed by a characteristic time, which differs from the
corresponding  characteristic time for the stationary potential
$V(x)=V_{f}(x)$. We note that the characteristic time is the same
as Kramers' time where the potential $V_{f}(x)$ is one with a
barrier .

 Our main results are given in the next Section, where we
derive the time--dependent probability distribution for a specific
form of the time variation leading from the initial potential
$V_i(x)$ to the final $V_f(x)$. Other forms of the time variation
can be treated by a simple extension of the techniques outlined
there. In Section III,  two specific examples of evolving
potentials are considered. When the time--scale of the
perturbation matches the Kramers time for the stationary potential
$V_{f}(x)$, there is a {\it resonance}  which delays the onset of
equilibrium.   This case is treated within the time--dependent
perturbation theoretic method of Dirac.  The second case we study
is one where $V_{i}(x)$ has a single minimum while $V_{f}(x)$ is
bistable.  By reducing to an effective  two--state dynamics, we
show that the Kramers time for the stationary potential $V_{f}(x)$
is reduced.   The paper concludes with a summary and discussion in
Section IV. Our result also sheds some light on the global
optimization scheme  recently introduced by  Hunjan, Sarkar and
Ramaswamy (HSR) \cite{21,hmsr}.

\section{The time--dependent distribution}
For concreteness, we consider the time--dependent potential

\bea
V(x,t)=V_{f}(x)+[V_{i}(x)-V_{f}(x)]e^{-\lambda t}
\eea

which evolves via homotopy from $V_{i}(x)$ at $t=0$ 
to $V_{f}(x)$ at $t \to \infty$.   The
Fokker--Planck equation for the probability distribution
$P(x,t)$ is

\bea\label{eq3} 
\frac{\partial P}{\partial t}=\frac{\partial}{\partial
x}\Bigg(P\frac{\partial V}{\partial x}\Bigg)+\epsilon
\frac{\partial^{2}P}{\partial x^{2}},
\eea

which, with the  substitution \cite{23,24,25},

\bea
\label{1}
P(x,t)=\phi (x,t)
\exp{-\frac{V(x,t)}{2\epsilon}},
\eea
 reduces to

\bea
\label{2}
\frac{\partial\phi}{\partial t}= H_{0}\phi+ H_{1}(t) e^{-\lambda t} \phi,
\eea
where (primes denoting differentiation with respect to $x$),
\ber
H_{0}&=& \epsilon\frac{\partial^{2} }{\partial
x^{2}}+\frac{1}{2}V_{f}^{\prime \prime}
-\frac{1}{4\epsilon}{V_{f}^{\prime}}^{2},\\
H_{1} &=& \Bigg[\frac {\Delta V^{\prime \prime}}
{2}-\frac{\lambda}{2\epsilon}\Delta
 V-V_{f}^{\prime}\frac{{\Delta V}^{\prime}}{2\epsilon}\Bigg]
-\frac{({\Delta V^{\prime}})^{2}}{4\epsilon}e^{-\lambda t}\\
\mathrm{and}\nonumber\\
\Delta V &=&V_{i}(x)-V_{f}(x).
\eer

$H_{0}$ satisfies the eigenvalue equation

\bea
H_{0}\psi_{n}= - E_{n}\psi_{n}(x)
\eea

with nonpositive eigenvalues. By construction, 
the ground state  has eigenvalue $E_{0}= 0$,  
with the eigenfunction  
$\psi_{0}(x) = A \exp  -V_{f}(x)/{2\epsilon}$, $A$ 
being a normalization constant.   Denoting the 
space--independent part of  $H_{1}$ by $V_{0}(t)$, 
the  solution of Eq.~(\ref{eq3}) can be written as

\bea
\label{eq7}
\phi(x,t)=\sum c_{n}(t)e^{-E_{n}(t)+
\int V_{0}(t^{\prime})dt^{\prime}} \psi_{n}(x).
\eea

Application of the standard  techniques of the Dirac
time--dependent perturbation theory  leads to 
\bea \label{eq8}
\dot{c}_{n}(t)=\sum_m  c_{m}(t)\langle m\vert H_{2}\vert n \rangle
e^{-(E_{m}-E_{n})t} \eea where \bea \label{eq8a} H_{2}=e^{-\lambda
t} (H_{1} - V_{0}). \eea The perturbative expansion for the
coefficients in Eq.~(\ref{eq7}) is in powers of  $H_{2}$ \bea
\label{eq9} c_{n}(t) = \sum_{j=0}^{\infty} c_{nj}(t) 
\eea

with

\bea 
\dot c_{n0}=0
\eea 
and for $j\ge$ 1,

\bea 
\label{eq11} \dot c_{nj}(t)=\sum_m
c_{m,j-1}(t)\langle m\vert H_{2}\vert n\rangle e^{-(E_{m}-E_{n})t}.
\eea

It can be seen immediately that the $c_{n0}$'s are  
constants determined by the state of the system at $t=0$.  
Assuming that the system is in the equilibrium state of the 
potential $V_{i}(x)$ at $t=0$,  namely

\bea
 P(x,0) = A_{0}\exp{-\frac{V_{i}}{\epsilon}},
\eea

this implies that

\bea
\phi(x,0) = P(x,0) \exp{\frac{V_{i}}{2\epsilon}}  = 
A_{0}\exp{-\frac{V_{i}}{2\epsilon}}
\eea
with normalization constant $A_{0}$.   
The constants $c_{n0}$ are now determined from the 
intial condition, as

\bea
\label{eq13}
c_{n0}=A_{0}\int_{-\infty}^{\infty}dx \psi_{n}(x) 
\exp{-\frac{V_{i}}{2\epsilon}}.
\eea

Substituting this in Eq.~({\ref{eq11}),  the 
complete solution to the problem can be obtained 
using Eqs.~(\ref{eq9}) and (\ref{eq7}).

Note that as $t\rightarrow \infty$, $\phi (x,t)$ 
in Eq.~(\ref{eq7}) tends to
$c_{00} \psi_{0}(x) =c_{00}\exp{-\frac{V_{f}}{2\epsilon}}$, 
and therefore,

\bea
P(x,t \rightarrow \infty) \rightarrow A \exp {-\frac{V_{f}}{\epsilon}},
\eea

the equilibrium distribution corresponding to $V_{f}(x)$.

Also note from Eqs.~(\ref{eq8}-\ref{eq8a}) 
that since $H_{2}$ has the time dependence $\exp{-\lambda t}$, 
there will be a resonance when $E_{m} - E_{n}=\lambda$, 
giving a secular growth of the first order term,  
$c_{n1}(t) \propto t$. This is analogus to case of the 
time--dependent perturbation theory \cite{26,27} 
in quantum mechanics. 

\section{Applications}

\subsection{Case I}
Consider first a case where the initial and final potential have the
same number of minima.   Specifically, we take $V_{i}=(x-a)^{2}$ and
$V_{f}(x)=x^{2}$, namely harmonic potentials that are spatially displaced. This leads to

\bea
\Delta V = a^{2}-2ax,
\eea

\bea 
H_{1} = \Bigg[-\frac{\lambda
(a^{2}-2ax)}{2\epsilon} + \frac{2ax}{\epsilon} \Bigg] -
\frac{a^{2}e^{-\lambda t}}{\epsilon},
\eea

which has the space--independent part
\bea 
V_{0}(t) =
-\frac{\lambda a^{2}}{2\epsilon} - \frac{a^{2}
e^{-\lambda t}}{\epsilon},
\eea

giving

\bea
H_{2}(x,t) = \frac{a x}{\epsilon}(\lambda + 2)e^{-\lambda t}.
\eea

The leading term in the expansion, namely
\bea
H_{0} = \epsilon \frac{\partial^{2}}{\partial x^{2}} + 1 -
\frac{x^{2}}{\epsilon}
\eea

has the eigenvalue spectrum  $E_{n} = 2n (n = 0,1,2, \ldots)$ 
with eigenfunctions

\bea
 \psi_{n}(x)  = \Bigg[2^{n} n!\sqrt{\epsilon \pi}
\Bigg ]^{-1/2} \H_{n}(x/\sqrt{\epsilon})\exp{-\frac{x^{2}}{2\epsilon }}
\eea

where $\H_{n}(y)$ are the usual Hermite polynomials.   
The time--dependent probability is therefore

\bea
P(x,t) = \phi(x,t)\exp(-\frac{x^{2}}{2\epsilon} -
\frac{(a^{2}-2 a x)}{2\epsilon}e^{-\lambda t})
\eea

giving, for  $\phi(x,t)$, the expansion (cf. Eq.~(\ref{eq7}))

\bea
 \phi(x,t) = \sum_n c_{n}(t)\exp(-2nt - \frac{a^{2}}
{2\epsilon}(1-e^{-\lambda t}))
\psi_{n}(x).
\eea

To first--order in the perturbation expansion, we find

\bea
\label{eq21}
 c_{n}(t=0) = \Bigg[\frac{a}{2\epsilon^{\frac{1}{2}}}\Bigg]^{n}
\frac{e^{-\frac{a^{2}}{2\epsilon}}}{[2^{n}n!]^{\frac{1}{2}}}.
\eea

At the lowest order of perturbation theory, only 
the $c_{n0}$'s, which are given by Eq.~(\ref{eq21}) matter.   
Straightforward algebra now shows that

\ber
P(x,t) &=& \Bigg[\frac{1}{\pi \epsilon}\Bigg]^{\frac{1}{2}}
\exp -\frac{x^2+a^2}{\epsilon}
+ \frac{2a x e^{-2t}}{\epsilon}  \nonumber\\
& & \exp \frac{a x (e^{-\lambda t})-e^{-2t}}
{\epsilon}+ \textit{O}(c_{n1}).
\eer

Since all terms to first order have not been 
included in the perturbation,
the equilibrium distribution is not properly
normalized as $t \to \infty$ and has the extraneous factor
$\exp(-{a^{2}}/{\epsilon})$. After computation of  
$c_{n1}$  from Eq.~(\ref{eq11}) we find, after 
taking the appropriate matrix elements and carrying out 
the integration over time, that

 \ber
 c_{n1}(t) &=& \frac{a}{\epsilon} \Bigg( \frac{n+1}{2} 
\Bigg )^{\frac {1}{2}}
\Bigg[ 1-e^{-\lambda t} \Bigg ] c_{n+1}(0)\nonumber\\
&+& \Bigg( \frac{\lambda +2}{\lambda -2}\Bigg ) \frac{a}{\epsilon}
\Bigg ( \frac{n}{2} \Bigg )^{\frac{1}{2}} \Bigg [1-e^{-(\lambda
-2) t}\Bigg ] c_{n-1}(0)
\eer

The coefficients $c_{n \pm 1}(0)$ are known from
Eq.~(\ref{eq21}) and after some amount of algebra we get, correct to
first order in the perturbing `Hamiltonian',

\bea 
P(x,t)=\Bigg (
\frac{1}{\pi \epsilon}\Bigg )^{\frac{1}{2}} e^{-{x^{2}}/{\epsilon}}
\Bigg [ 1 + \frac{2 a x e^{-\lambda t}}{\epsilon} + \frac{4 a x
(e^{-2t}-e^{-\lambda t})}{(\lambda -2)\epsilon} +
\ldots \Bigg ]
\label{result}
\eea

The effect of time--dependence in the potential can 
be seen by contrasting the above result, Eq.~(\ref{result}) 
with the sudden limit, when the potential is 
instantaneously changed from $V_i(x)$ to  $V_{f}(x)$. 
The initial probability distribution corresponding to 
$V_i(x)$ is  $\Bigg (\frac{1}{\pi \epsilon}\Bigg)^
{\frac{1}{2}}e^{-\frac{(x-a)^{2}}{\epsilon}}$ and 
this approaches the equilibrium distribution 
corresponding to $V_f(x)$  as

\bea
P(x,t) = \Bigg (\frac{1}{\pi \epsilon}\Bigg )^{\frac{1}{2}} 
e^{-\frac{x^{2}}{\epsilon}} \Bigg [ 1 + \frac{2 a x
e^{-2t}}{\epsilon}\Bigg ].
\eea

The coefficient $\frac{2 a x}{\epsilon}$ of $e^{-2t}$ 
has the extra factor $(1+{2}/(\lambda -2))$.  Note 
that the time--dependent perturbation effectively keeps 
the system from attaining equilibrium by always managing 
to cause transition to neighbouring states. The approach 
to equilibrium depends on the  value of the 
adiabaticity parameter, $\lambda$, and there are three 
different regimes of interest.

\begin{itemize}
\item
When  $\lambda > 2$,  the approach to equilibrium is 
governed by $e^{-2 t}$ but the coefficient of this term 
is significantly increased.
\item If $\lambda < 2$, the approach is controlled by 
$e^{-\lambda t}$ and in the
long time limit

\bea
P(x,t)\sim \frac{1}{({{\pi \epsilon}})^{1/2}}
e^{-{x^{2}}/ \epsilon}\Bigg [ 1 + 
\Bigg (\frac {4 a x/\epsilon}{2-\lambda
}\Bigg ) e^{-\lambda t} \Bigg ].
\eea
\item

Finally, we have the extremely interesting situation of 
$\lambda \simeq 2$,
in which case

\bea
P(x,t) = \frac{1}{(\pi \epsilon)^{1/2}} 
e^{-\frac{x^{2}}{\epsilon}} \Bigg
( 1 + \frac{4 a x t}{\epsilon}e^{-2t} + \ldots \Bigg )
\eea

This is the resonance that we have discussed 
already, which shows up as the coefficient of the
usual correction to $P_{eq}(x)$ diverging with time.

This divergence of the coffecient of $e^{-2t}$ in Eq.(34)
would eventually get transferred to the argument of the 
exponential function as is usual in such cases. This 
can be explicitly verified in this case, because an exact 
solution for harmonic potentials has been written down 
by H\"{a}nggi and Thomas \cite{28}. The answer for $P(x,t)$, adapting
the work of H\"{a}nggi and Thomas to this situation is 
  
\bea
 P(x,t)= \Bigg [\frac{1}{\pi \epsilon 
(1-e^{-4t})}\Bigg ]^{1/2} 
\exp \Bigg [-\frac{\Big (x-ae^{-2t}[1+2(1-e^{-(\lambda -2)t})/(\lambda -2)]
\Big )^{2}}{\epsilon (1-e^{-4t})}\Bigg ]
\eea

If we expand the exponential in powers of $'a'$ all the three
cases cited above are exactly reproduced. This shows that the method
of quantum mechanical time dependent perturbation theory
that we have adopted here in capable of yielding the correct results. 

\end{itemize}

\subsection{Case II}

We now turn to a situation where an initially single well
structure, $V_{i}=x^{2}$,  crosses over to a double well
structure, $V_{f}=-\frac{x^{2}}{2}+\frac{x^{4}}{4}$,  as $t \to
\infty$. The approach to equilibrium in double well potential is
governed by the Kramers time, the long time scale coming from the
possibility of noise induced hopping .  Following the procedure
outlined in Eqs.~(\ref{1}-\ref{2}), we get 

\ber
H_0&=& -\epsilon \frac{d^2}{dx^2}+\frac{(x^3-x)^2}
{4\epsilon}-\frac{1+3x^2}{2}\\
\Hp&=& \frac{3}{2}
(1-x^{2}) -\frac{1}{4\epsilon} x^{2} (3-x^{2})
(1-x^{2})\nonumber\\
&-& \frac{\lambda}{8\epsilon} x^{2} (6-x^{2}) +\frac{1}{4\epsilon}
x^{2} (3-x^{2})^{2} (1-e^{-\lambda t}).
\eer

The low--lying part of the eigenvalue spectrum of 
$H_{0}$ is characterized by
a set of close doublets with exponentially small 
separations, while the gap between two
doublets is of \textit{O}(1).   The ground 
state $E_{0}=0$, while the first excited state 
is the ground state of the supersymmetric partner of

\bea
\frac{(V_{f}^{\prime})^{2}}{4\epsilon}-\frac{V_{f}^{\prime \prime
}}{2}
\eea

and is exponentially small \cite{29,30}.   
The next excited  state has eigenvalue approximately  2,  
and hence we can treat  the dynamics of the low lying 
states as that of a two level system.
Denoting the two states by $\phi_{0}$ and $\phi _{1}$, with 
eigenvalues 0  and
$\delta$, then

\bea
\label{eq29}
\phi (x,t) =c_{0}(t)
\phi _{0} (x) + c_{1}(t) e^{-\delta t} \phi _{1}(x)
\eea

The dynamics of $c_{0}$ and $c_{1}$ is governed by

\bea
\dot c_{0} =
\langle\phi _{0}\vert \Hp\vert \phi_0\rangle 
e^{-\lambda t} c_{0}(t) + \langle\phi
_{0}\vert \Hp\vert\phi _{1}\rangle  c_{1}(t) e^{-(\lambda +
\delta)t}
\eea

\bea\label{eq31} 
\dot c_{1} = \langle\phi_{1}\vert \Hp\vert
\phi_{0}\rangle  e^{-(\lambda -\delta)t} c_{0}(t) + 
\langle\phi_{1}\vert
\Hp\vert \phi_{1}\rangle e^{-\lambda t} c_{1}(t)
\eea

Since the perturbation $\Hp$ is even,  $\langle\phi_{1}\vert
\Hp\vert \phi_{0}\rangle =0$, decoupling  $c_{0}$ and $c_{1}$
Integrating Eq.~(\ref{eq31}) and dropping terms
like $e^{-2 \lambda t}$ which are unimportant for $t>\lambda
^{-1}$, we find

\bea 
\dot c_{1} = c_{1} \langle\phi_{1}\vert \widetilde
{\Hp}\vert \phi_{1}\rangle  e^{-\lambda t}
\eea

where
\bea
\widetilde {\Hp} = \frac{3}{2} (1-x^{2}) +
\frac{1}{2\epsilon } x^{2}(3-x^{2})(1-x^{2}) - \frac{\lambda
}{8\epsilon } x^{2} (6-x^{2})
\eea

The primary contribution to
$\langle\phi_{1}\vert \widetilde {\Hp}\vert \phi_{1}\rangle $  comes
from the vicinity of $x=1$ since $\phi_{1}$ is an antisymmetric
wave function strongly peaked near $x\pm 1$.   Since
$\langle\phi_{1}\vert \widetilde {\Hp}\vert \phi_{1}\rangle  \simeq -
\frac{5\lambda}{4\epsilon}$, we have

\bea 
c_{1}(t) = B \frac{1-e^{-\lambda t}}{\lambda} 
\exp{-\frac{5 \lambda}{4\epsilon}}
\eea

with $B$ a constant of integration.   From
Eq.~(\ref{eq29})  we  find, after a series of standard
manipulations, that
 
\ber 
\phi (x,t) &\simeq & N \phi_{0}(x) \Bigg [ 1 +
\frac{\phi_{1}(x)}{\phi_{0}(x)} e^{-\delta t -\frac
{3(1-e^{-\lambda t})\alpha}{2\lambda}}\Bigg ]\nonumber\\
&\equiv& N \phi_{0}(x) \Bigg [ 1 + f(x) e^{-\delta_{\eff} t} \Bigg].
\eer

In the above $\alpha$ is a measure of the strength of the ground state
wave function at the origin, and
where

\bea 
\delta_{\eff} = \delta + \frac{3(1- e^{-\lambda t})\alpha}{2 \lambda t}
\eea

The inverse of $\delta_{\eff}$
gives the effective Kramers time for the system and is shorter
than the scale for the time independent system.   This speeding up
is most effect in the adiabatic limit, namely for $\lambda \ll 1$.

\section{Discussion and Summary}

The above result is a simple analogue of the global optimisation
principle on an evolving energy landscape.   In this case one is
interested in finding the minima of a multidimensional potential
energy surface which constitutes the energy landscape in problems such
as protein folding or finding the ground state configuration of atomic
or molecular clusters.   The observation of  HSR \cite{21,hmsr},  that
continuously and adiabatically varying potentials assist approach to  
the desired
configuration at $t\rightarrow \infty$ by avoiding trapping in local
minima.   We have shown in a model system,
a similar time dependence , the decrease of Kramers time makes escape
from a trapping potential easier.

The above demonstration of a reduced Kramers' time is for a 
one dimensional system. The extension to two dimensions is 
reasonably straightforward following the technique in {\it ref.}[30]
when a well defined tunneling path exists between the two minima.
Extensions to more general situations and to three dimensions is being
investigated. We have also seen that this technique of dealing with
time dependent perturbations can model the stochastic resonance
as a kind of parametric resonance. With the emerging importance
of stochastic resonance in biological systems \cite{10} it is 
possible that yet another way of  looking at stochastic
resonance can yield new insights.

\end{document}